\documentclass[preprint,10pt]{sigplanconf}

\pdfoutput=1

\usepackage{amsmath}
\usepackage{listings}
\usepackage{graphicx}
\usepackage{hyperref}

\usepackage[T1]{fontenc}
\usepackage{inconsolata}

\begin{document}

\lstset{language=Java, frame=none, showstringspaces=false, basicstyle=\footnotesize\ttfamily, tabsize=2, breaklines=true}
\newcommand{\code}{\texttt}

\conferenceinfo{WXYZ '05}{date, City.} 
\copyrightyear{2005} 
\copyrightdata{[to be supplied]} 


\title{Poplar: A Java Extension for Evolvable Component Integration}

\authorinfo{Johan Nystr\"om-Persson \and Shinichi Honiden}
           {University of Tokyo and National Institute of Informatics, Tokyo, Japan}
           {jtnystrom@is.s.u-tokyo.ac.jp \and honiden@nii.ac.jp}

\maketitle

\code{\hyphenchar\font=-1}

\begin{abstract}
The Java programming language contains many features that aid component-based software development (CBSD), such as interfaces, visibility levels, and strong support for encapsulation. However, component evolution often causes so-called breaking changes, largely because of the rigidity of component interconnections in the form of explicit method calls and field accesses. We present a Java extension, Poplar, which we are currently developing. In Poplar, inter-component dependencies are expressed using declarative queries; concrete linking code, generated using a planning algorithm, replaces these at compile time. Poplar includes a minimal specification language based on typestate-like protocols and labels, and a lightweight effect system, which ensures the absence of unwanted interference between hand-written code and generated code. 
We give several examples of fully automatic component integration using Poplar, and demonstrate its potential to simplify object-oriented software development greatly through evolvable and statically checkable integration links. 
\end{abstract} 
\category{D3.3}{Programming Languages}{Frameworks}
\category{D3.3}{Programming Languages}{Constraints}

\terms
Languages

\keywords
CBSD, protocols, programming languages, code synthesis, typestate, object-oriented programming, planning, adaptation, evolution

\section{Introduction}

Two essential and related properties of object-oriented programming languages like Java are encapsulation and polymorphism. Encapsulation is the principle of separating interface from implementation, and this in turn enables polymorphism, whereby the runtime type of an object may be different from its declared type in the source code, and thus unknown to the caller. When two classes have the same interfaces, according to the principle of behavioural subtyping\cite{liskov}, the implementations should be substitutable for each other and all expected safety properties should be retained. Contemporary programming paradigms such as component-based software development (CBSD)~\cite{osgi, javabeans} draw heavily on these principles. 


Theoretically, interfaces of classes should only change in backward-compatible ways once they have been published, for instance through the addition of new methods, or through the widening of assumed preconditions and narrowing of assumed postconditions. Interface changes that require client classes to update their associated client code are called breaking changes. While developers strive not to make such breaking changes, it has been found that in practice they are commonplace~\cite{Dig2005}. Breaking changes introduce a large cost into component-based software development, since potentially every dependent class may have to be updated. In other words, CBSD, as it is practised today, suffers from a conflict between software evolution and flexibility of composition. The more likely a class is to evolve, the greater the potential future cost of having included it in a software system. Moreover, the need for class evolution is unlikely to go away, since few software specifications are final. Classes may receive new features, changes in the domain that they model may necessitate an implementation change, bugs may be fixed, and so on.

Tentatively, we can identify certain categories of language-level breaking changes that can occur in modern imperative object-oriented languages, such as Java, C\# and C++.

\begin{description}
\item[Name changes.] The renaming of a method, everything else being the same.
\item[Protocol changes.] Often, a sequence of method invocations is required to establish a certain effect or compute a certain value. When this sequence changes from one class version to another, we say that a protocol change has occurred. This includes permutations of protocol steps, but also addition of new steps and removal of old steps. Values may pass between method invocations in a protocol fragment in a complex way; changes in these inter-method dependencies with respect to a result should also be considered to be protocol changes.
\item[Type changes.] Methods may be moved from one class to another; argument and return types may be changed to incompatible types. 
\item[Signature changes.] The number of arguments that methods require may change, without visibly affecting the functionality that existing clients receive. 
\end{description}

In addition, there are changes that occur above the level of the language, such as conceptual semantic changes and quality attributes~\cite{Becker2006}.


 Considering these problems, we believe -- as did Shaw~\cite{Shaw1993}  in 1993 -- that traditional class interfaces, consisting of collections of named methods and their signatures, lead to a very rigid form of class interconnections, which permit little evolutionary flexibility. Current interfaces expose inessential information, such as method names and argument ordering, that clients become dependent on, while at the same time not exposing essential information, such as semantics, valid class protocols and valid interactions, that clients need in order to construct and verify integrating links. We are currently developing a Java extension, Poplar, which attempts to reinvent component integration to address these problems. Poplar adds several concepts to the Java language to support a new kind of composition methodology, in which integration requests are expressed using declarative \emph{queries}. At compile time, we generate concrete linking code, whose content will reflect factors such as what components are available and how they have evolved.

 
We face several essential problems that need to be addressed in order to make this approach work. 

\begin{description}
\item[Complexity of code synthesis] Code synthesis is generally a difficult problem. Whole-program synthesis is generally successful only for small examples, and even then is often very computationally expensive, due to the use of a theorem prover or a constraint solver~\cite{136556, oracleGuided}. Generally, a lot of this complexity stems from the use of a rich description language, such as Hoare logic or OCL. We believe that, since we are not attempting whole-program synthesis, our description language should be simplified as much as possible in order to reduce the complexity of the synthesis problem. Our first design principle is to use a \emph{minimal description language}. In designing this, we draw on the idea of typestate protocols~\cite{Strom1986,Deline}, which is a natural way of constraining and describing API usage in a useful, compact and evolvable manner. We also use atomic, uninterpreted labels, which have been used previously in the context of ML\cite{Haack2002}.

\item[Effects and interference] Since Java expressions may have side effects, it becomes necessary to reason about potential interference caused by these in some way. In general, we will be generating short Java code fragments that are to be inserted at the location of integration queries in client code, and thus mixed with hand-written code. For each generated statement, it is essential to know whether it may interfere with other generated statements, or with the pre-existing hand-written statements. In order to model this, we use a simple effect system based on \emph{abstract resources}, which are an adaptation of Boyland and Greenhouse's abstract regions~\cite{objOrientedEff}.

\item[Aliasing] Third, in order to reason on side effects on a per-object basis, we will need some way of constraining aliasing, since in Java, each object may potentially have any number of references. For this, we use a minimal uniqueness system~\cite{Minsky:1996fk}, where references and method arguments are classified according to what assumptions they make about references, and whether they may create any aliases.
\end{description}

In choosing what effect system and aliasing policy to use, the rationale has been to use minimal concepts that are easy to understand and discuss, in order to allow us to design a framework with relatively clear functional and validity properties. However, future improvements to the accuracy of the effect system or the aliasing policy should be independently possible without great impact to the overall design. Our framework also does not impose any particular constraint on what algorithm or algorithms to use in the generation of solutions, but a natural choice is Partial Order Planning (POP)~\cite{Mcallester1991}, which has shown promise in a very early prototype that we have tested. 

Our design permits both modular integration and modular checking of integrations.

We make the following contributions.
\begin{itemize}
\item We introduce Poplar, a Java extension that provides a framework for declarative specification of integration queries, automated code generation for their satisfaction, and effect protection.
\item We outline how the POP algorithm can be used to generate solutions to integration queries in our framework.
\end{itemize}

The rest of the paper gradually explains and illustrates features of Poplar using examples. We introduce the basic ideas of labels, queries and solutions (Section~\ref{sec:basic}), followed by type protocols and abstract resources (Section~\ref{sec:protocols}). In Section~\ref{sec:uniqueness}, we describe how effects may be protected using unshared references and effect spans. We introduce a realistic example based on the Swing GUI toolkit in Section~\ref{sec:interclass}, and show how subclasses may extend and override resources, protocols and annotations from superclasses. We also introduce interclass protocols, which describe how classes can interact in useful ways. In Section~\ref{sec:planning}, we discuss the details of code generation using POP. We show how Poplar can be adopted and used in practice, as well as discuss its limitations in Section~\ref{sec:discussion}. Finally, we discuss related work (Section~\ref{sec:related}) and conclude the paper (Section~\ref{sec:conclusion}).

Throughout this paper, we disregard issues raised by concurrency or reflection. We discuss a possible approach to exceptions in Section~\ref{sec:conclusion}.
\section{A basic example}
\label{sec:basic}

We introduce the Poplar Java integration mechanism using a simple example from the real world. The time and date API of the standard Java libraries changed substantially between version 1.4 and version 1.5 of the language. In version 1.4, the following code was used to obtain the current hour of the day:

\begin{lstlisting}
Date now = new Date();
int hour = now.getHour();
\end{lstlisting}

In Java 1.5 and later versions, the following code is used:

\begin{lstlisting}
Calendar now = Calendar.getCalendar();
int hour = now.get(Calendar.HOUR_OF_DAY);
\end{lstlisting}

Even though the Java 1.5 libraries keep the old version of the API, this is representative of a breaking change that may occur in practice, and API publishers generally prefer not to have to preserve old versions.

In principle, Poplar considers integration sites to have one of two possible purposes: producing values or producing effects. In these two cases we use produce queries and transform queries, respectively. Clearly, in this case, a client component that wants to use the time and date API wants to do the former. Before we can request the production of a value, we must annotate the API that is provided by the service component. In the case of Java 1.4, the component supplier should provide annotations similar to the ones in Figure~\ref{fig:tad14}.

\begin{figure}
\begin{lstlisting}
interface TimeAndDate {
	labels(int) nowHour, nowMinute, nowSecond; 
}

class Date implements TimeAndDate {
	labels currentTime;
	Date()
		result: +currentTime;
	int GetHour()
		this: currentTime,
		result: + nowHour;
	/* Similar annotations for getMinute(), getSecond(), etc. */
}
\end{lstlisting}
\caption{\code{TimeAndDate} annotations for Java 1.4.}
\label{fig:tad14}
\end{figure}

\subsection{Labels}
We have added the \code{labels} annotation as a new member of classes and interfaces. In the case of the interface \code{TimeAndDate}, labels are provided for the int type using the notation \code{labels(int)}. Once these labels have been defined, we can logically distinguish between integers that have these labels and integers that do not, as a lightweight refinement of the type system. Since the \code{Date} class implements \code{TimeAndDate}, references to \code{nowHour} in this class are understood to refer to the label defined in the \code{TimeAndDate} interface, but it is possible for other interfaces to define labels with the same name, and their meaning might be different. Disambiguation should be done in the usual way using fully qualified names where necessary.

In the \code{Date} class we have added pre- and postconditions to the methods. The constructor \code{Date()} declares that the result, ie the return value, will have the new label \code{currentTime}. This label was declared in the \code{Date} class itself. The + sign indicates that a new label is added. In contrast, the \code{getHour} method indicates that for the this variable, the receiver of the method, an invariant of the \code{currentTime} label is expected - the label must be owned by the this object prior to method invocation, and it will remain after the invocation. When this method is invoked, the return  value will be an integer which has the \code{nowHour} label. Here, the labels describe one kind of useful application of the method, but not mandatory constraints on it. It is still valid to invoke this method when the this variable does not have the \code{currentTime} label, but in this case, unless the annotations are augmented beyond what is shown here, we can make no assumptions about the return value. 
As for the client component which wants to produce the value corresponding to the current hour of the day, its code should resemble the following:

\begin{lstlisting}
class TimeUtils implements TimeAndDate {
	void printHour() {
		int hour = #produce(int, nowHour);
		System.out.println("The current hour is: " + hour);
	}
}
\end{lstlisting}

Again, we use the \code{TimeAndDate} interface to refer to the \code{nowHour} label. We request a production of an integer value with this label using the \code{\#produce} query. We prefix queries with a '\#' sign to distinguish them from normal Java code. At compile time, the Poplar solver will find a solution to this query and replace it with a sequence of Java statements. Such statements can be field accesses or method invocations, including constructor invocations. In principle, queries may be inserted anywhere that a sequence of statements may occur in Java. Code resulting from queries may return a single value, and the queries may thus be ``assigned'' to a variable, as in the example we have just shown.

The solver uses a planning algorithm to find a solution to the query. The plan search will proceed backwards from the goal to the assumptions. In this case, the goal is the existence of a variable of type \code{int} and with label \code{nowHour}. First the planner needs to find all actions that can produce such a variable (method invocations and field accesses). If the only one available is the one we declared above (\code{Date.getHour}), then once this method has been selected, a new set of preconditions will result - in order for that method to be invoked, we need to have a \code{Date} object with the \code{currentTime} label. We repeat the search and find that there is a constructor that takes no arguments and that produces such an object. This yields a complete solution, and thus, after the code has been generated and the substitution has taken place, the client class will look like the following:

\begin{lstlisting}
class TimeUtils implements TimeAndDate {
	void printHour() {
		Date v1 = new Date();
		int hour = v1.getHour();
		System.out.println("The current hour is: " + hour);
	}
}
\end{lstlisting}

In addition, the solver will remove non-Java elements from the code, such as the label declarations from the various classes, so that the result is valid Java source code.

\subsection{Upgrading to Java 1.5}

Let us now consider how we could adapt this client to the Java 1.5 version. In this case, the service component would resemble the one shown in Figure~\ref{fig:tad15}.

\begin{figure}
\begin{lstlisting}
Class Calendar implements TimeAndDate {
	labels(int) hourMarker, minuteMarker, secondMarker;
	labels defaultTimeZone;

	final int HOUR_OF_DAY +hourMarker = 11; 
	/* etc. */

	Calendar()
		result: +defaultTimeZone { ... }
	int get(int selector)
		this: defaultTimeZone,
		(selector: hourMarker,
		result: +nowHour)?,
		(selector: minuteMarker,
		result: +nowMinute)? { ... }
}
\end{lstlisting}
\caption{\code{TimeAndDate} annotations for Java 1.5. Here the + sign indicates an invariant.}
\label{fig:tad15}
\end{figure}

\begin{figure}
\begin{center}
\includegraphics[scale=1.18]{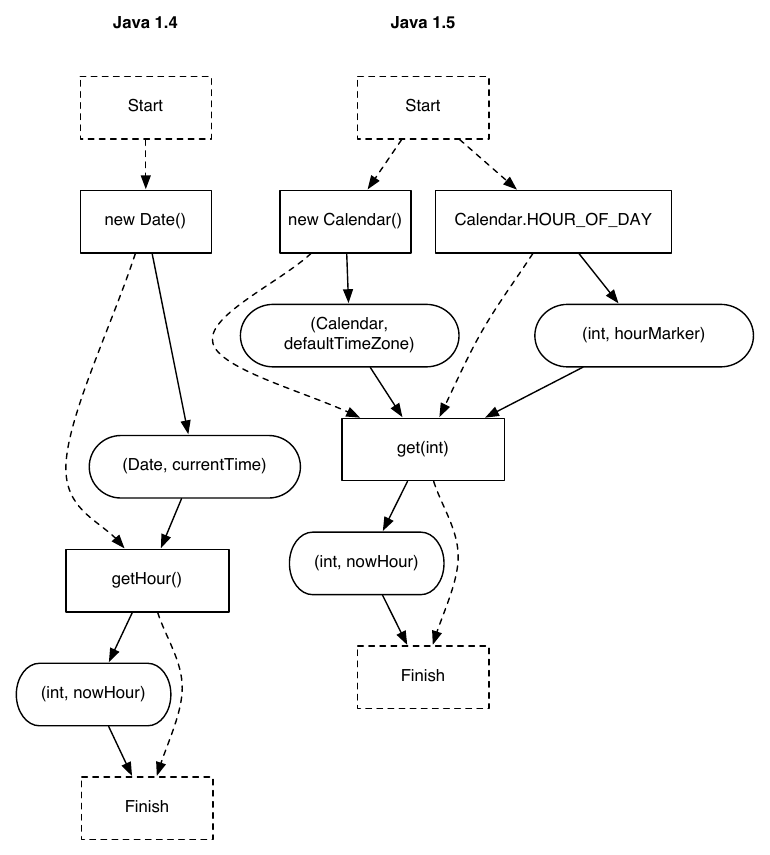}
\caption{Visual representations of the plans generated in the case of producing the current hour of the day in Java 1.4 and 1.5, respectively. }
\label{fig:date}
\end{center}
\end{figure}

	In this case, all the values are accessed through one method, which takes a selector argument. We have given the field \code{HOUR\_OF\_DAY} an explicit label \code{hourMarker}, which links it to its possible use as an argument for the get(int) method. We group invariants and postconditions using the \code{(a, b, \ldots)?} syntax, which makes the pre- and postconditions inside the group optional. This is mainly a syntactic shorthand to avoid having to repeat the \code{defaultTimeZone} requirement for the this object. In this case, however, clearly \code{minuteMarker}, \code{hourMarker} and so on will be mutually exclusive labels. The plan search now takes the same query as a starting point, but the APIs supplied as input are different (and perhaps the 1.5 API is marked as taking precedence over the 1.4 one if both are available) and after substituting a solution for the query we end up with:
	
\begin{lstlisting}
class TimeUtils implements TimeAndDate {
	void printHour() {
		Date v1 = new Calendar();
		int v2 = Calendar.HOUR_OF_DAY;
		int hour = v1.get(v2);
		System.out.println("The current hour is: " + hour);
	}
}
\end{lstlisting}

We show the plans visually in Figure~\ref{fig:date}. The resulting code is extracted directly from the plans. Rounded boxes are pre- and postconditions (the existence of a variable with a given label), and square boxes are actions such as method invocation and field access. Dashed lines represent sequential constraints, which impose an ordering on the actions. These constraints will be at least as strong as the dataflow dependencies of the solution, and possibly stronger due to possible conflicts. 

These examples demonstrate how clients can automatically be reconfigured to use a new version of an API, or even a different API, given that the necessary annotations are present on both the client and the service side. In this case it would simply be a matter of re-running the integration tool with the newest service components added to the classpath.

\subsection{Syntax}

We give the syntax of Poplar in Figure~\ref{fig:syntax}. We have omitted many inessential features, including constructors, imports and the interface forms of method declarations, as well as elements that are unchanged from standard Java, for brevity.

\begin{figure}
\begin{tabular*}{0.3 \textwidth}{@{\extracolsep{\fill}}llr}
$PR$			& ($\overline{CL}$, e)															& program \\		
$CL$			& \code{class} $C$ \code{extends} $C'$	 \{ $\overline{F}$ $\overline{R}$ $\overline{P}$ $\overline{M}$ $\overline{L}$ \} & class decl. \\
$F$			& [ u ] [ \code{managed} [($r$)] ] $T$ $f$ [ + $\overline{l}$ ] = $e$	& field decl. \\
$P$			& \code{protocols} $\overline{p}$	& protocol decl. \\
$R$			& \code{resource} $\overline{rd}$	& res. decl. \\
$rd$ 		& $r$ [ \{ $\overline{rd}$ \} ]			& res. def. \\
$qr$			& $r$ [.$qr$] 						& qualified res. \\
$L$ 		& \code{labels} [($\overline{T}$)] $\overline{l}$ & label decl. \\
$u$ 			& $\code{maintain} | \code{maintainr} | $ \\ 
			& $\code{unique} | \code{uniquer}$ & uniqueness  \\
			\\
$M$		& $md$ \{ $\overline{e}$ \}				& method def.	\\		
$md$ 		& [ $u$ ] [ + $\overline{l}$ ] $T$ $m$ ( $\overline{ma}$ ) '[' ! $\overline{qr}$ ']' \\
		& [$ms$ :] [$\overline{cj}$] [$\overline{(\overline{cj})?}$] 	& method decl.\\  
$ma$	& [$u$] $T$ $x$							& method arg. \\
$ms$	& \code{mutates} $\overline{\code{this.}qr | x.qr | x} $ 	& mut. summary \\
$cj$		& ($\code{this}|\code{result}|x$): $cd \overline{[, cd]}$	& conjunction \\
$cd$		& $i | ec$										& condition \\
$i$		& $l | p@s$										& invariant \\
$ec$		& $p@s \rightarrow s | +l$ [ '[' $* \overline{qr}$ ']' ]		& effect \\
\\
$ex$		& \code{external} $C$. $md$					& external decl. \\
$e$		& $je | q$									& expression \\
$q$		& \code{\#produce($T$, $l$)} $|$ \\
		& \code{\#transform($x$, $l$)} [ \{ $\overline{e}$ \} ] & query \\

\end{tabular*}
\vspace{1em}

$x$ variable, $T$ type, $C$ class, $f$ field, $l$ label, $r$ resource,
$m$ method, $p$ protocol, $s$ protocol state, $je$ Java expression
\caption{Syntax of Poplar. Elements that are unchanged from standard Java have been omitted. A horizontal bar $\overline{x}$ indicates repetition.}
\label{fig:syntax}
\end{figure}

\section{Protocols, resources and protection}
\label{sec:protocols}

Having explained basic synthesis based on labelled variables, we will now introduce the additional features of type protocols, abstract resources and effect protection. These features are intended to provide a basic level of safety and guarantee that generated code does not interfere with itself or with handwritten code.

\subsection{Type protocols}

Type protocols is a version of the technique known as typestate checking~\cite{Strom1986}. In typestate checking for objects~\cite{Deline}, every class is considered to possess a formal protocol described by a state machine, which regulates valid and invalid message sequences for that class. In Poplar, we do not check that existing code conforms to typestate protocols, but use these protocols as a generative constraint. We simply never generate any code that violates any declared typestate protocols. 

A network socket is a natural example. Sockets go through a well defined sequence of method calls during their lifetime: they cannot open a connection before they are bound, and they cannot send or receive data until they have been opened. When they have been closed they can no longer send or receive data. A partial socket class is shown in Figure~\ref{fig:socket}. An annotation like \mbox{\code{type@raw $\rightarrow$ bound}} specifies that the this object must be in the \code{type@raw} state prior to the method invocation, and will be in \code{type@bound} after the invocation. The benefit of this is that, for instance, if we request the production of an open socket from scratch, it must necessarily pass through all the intermediate steps first. We permit multiple protocols with different names in each class.

\begin{figure}
\begin{lstlisting}
class Socket {
	protocols type;
	Socket()
		result: +type@raw {...}
	void Bind(SocketAddress bindPoint)
		this: type@raw->bound {...}
	void connect(SocketAddress endPoint)
		this: type@bound->open {...}
	void Send(byte[] data)
		this: type@open {...}
	int receive(byte[]data, int offset, int max)
		this: type@open {...}
	void close()
		this: type@open->closed {...}
}
\end{lstlisting}
\caption{A socket class. Note the \code{type@a$\rightarrow$b} protocol annotations.}
\label{fig:socket}
\end{figure}

\subsection{Abstract resources}
Sometimes it is desirable to partially constrain the behaviour of a class, when its state can be naturally partitioned into distinct areas or modes of operation. For instance, a GUI widget that contains other widgets could have its state separated according to appearance and contents. In the case of the socket, we can separate its state according to the state of the connection and the data being sent. We do this using the syntax shown in Figure~\ref{fig:socket2}.

\begin{figure}
\begin{lstlisting}
class Socket {
	protocols type;
	resources connState, data;
	Socket()
		result: +type@raw {...}
	void bind(SocketAddress bindPoint)
		[!connState]
		this: type@raw->bound [*connState] {...}
	void connect(SocketAddress endPoint)
		[!connState]
		this: type@bound->open [*connState] {...}
	void send(byte[] data) [!data]
		this: type@open {...}
	int receive(byte[] data, int offset, int max) [!data]
		this: type@open {...}
	close()	[!connState]
		this: type@open->closed {...}
}
\end{lstlisting}
\caption{The socket class with resource mutations \code{([!x])} and residence \code([*x]) included.}
\label{fig:socket2}
\end{figure}

Here we have declared two \emph{abstract resources}, \code{connState} and \code{data}. Like labels and protocols, they can be declared in any classes or interfaces. A resource is a stateful asset of any kind, which may include Java fields and data accessed using JNI including external data such as sockets and graphics. Resources may be mutated, which logically means that some kind of change has occurred. An annotation such as \code{[!connState]} simply means that the \code{connState} resource is mutated if the corresponding method is invoked. Thus, another way of thinking about abstract resources is that they are a group of methods which may mutate a single property in some way. A resource \code{r} is defined by the set of \code{[!r]} annotations in the class that declares it. The exact nature of the resources is not identified beyond annotating the methods that may affect them. In this way, they differ from Boyland and Greenhouse's~\cite{objOrientedEff} abstract regions, which are similar to our resources, but that describe groupings of Java fields only, and thus cannot manage external state.

In addition to methods mutating abstract resources, achieved effects can \emph{reside} in them. This declares that the effect will last for as long as the resource is not mutated. Thus, for instance, the notation \code{[*connState]} of the \code{connect()} method indicates that the effect of transforming the state \code{type@bound} into \code{type@open} resides in the \code{connState} resource. The opened socket will remain in \code{type@open} as long as this resource is not mutated. An effect may reside in any number of resources. 


\subsection{Protection spans}
The benefit of this kind of annotation is to constrain the generated code further by indicating resources that are to be protected. For instance, consider the network server shown in Figure~\ref{fig:networkServer}. In this case, we are concerned about the socket being closed prematurely by mistake, and we would like to protect ourselves against this kind of mistake, whether from hand-written or generated code.

\begin{figure}
\begin{lstlisting}
void serveClient(...) {
	Socket s = #produce(Socket, type.open) 
	{ //Protection span begins here
	boolean quit = false;
		do {
			String r = readRequest(s);
			ClientRequest cr = #produce(ClientRequest, processed)
			quit = shouldQuit(cr);
		} while (quit == false);
	} //Protection span ends here
	#transform(s, type.closed);
}
\end{lstlisting}
\caption{A simple network server. Note the protection span, which is denoted by \code{\{ \}} signs, and emphasised by comments. Generated and hand-written code inside the span must not violate the protection rule.}
\label{fig:networkServer}
\end{figure}

The server repeatedly reads requests from the socket and processes them. Depending on the exact request, the \code{readRequest()} method, as well as the code generated in the \code{\#produce(ClientRequest, processed)} query, may dispatch the request to various different code locations, of which new ones may potentially be added in the future -- for example through polymorphism or through regeneration of the query's solution. Assuming that the produce query is resolved using the \code{Socket.open()} method given in Figure~\ref{fig:socket2}, we know that the effect \code{type.open} should reside in the resource \code{s.connState}, and thus it can be protected by ensuring that this resource is not mutated. In Figure~\ref{fig:networkServer}, we use standard curly braces to indicate the start and end of a protection span. Handwritten code inside the span is checked, and generated code is constrained for compliance. At the point of the \code{\#transform(s, closed)} query, the protection span has ended, and the socket can be closed in a valid manner. Depending on the precise way that the \code{\#produce(Socket, type.open)} query is satisfied, the protection span will have a different meaning, so if protection fails during code generation, the code generator may be spurred to attempt finding a different solution with more lenient or different resource protection needs. Note that a \code{\#transform} query is different from a \code{\#produce} query -- the former requests that a label be added to an existing value; the latter requests the production of a new value.

In addition to abstract resources belonging to instance variables, we may also define static abstract resources that correspond to invocation of static methods, similar to the approach taken in~\cite{objOrientedEff}.

Reasoning about side effects like this requires two new language elements that have not been introduced so far: uniqueness kinds and mutation summaries. We will discuss them in the following section.

\section{Mutation summaries and uniqueness}
\label{sec:uniqueness}

\subsection{Mutation summaries}
In order to protect resource such as \code{connState} in Figure~\ref{fig:networkServer}, we will need a way of gathering all the potential mutations that can occur as a result of invoking a method. We choose to annotate methods with \emph{mutation summaries}, in the style of Boyland and Greenhouse's effect summaries~\cite{objOrientedEff}. They simply summarise all the potential (recursive) effects of invoking a method. Assume the network socket introduced earlier, and consider the following utility method.

\begin{lstlisting}
void connectAndSend(maintain Socket s, SocketAddress sa, byte[] data) 
	mutates s.connState, s.data: {
		s.connect(sa);
		s.send(data);
}
\end{lstlisting}

In this case, since the \code{s.connect} and \code{s.send} methods mutate the \code{s.connState} and \code{s.data} resources, we have listed them in the mutation summary. 

Mutation summaries are simply lists of resources, protocols and \code{managed fields} (Section~\ref{sec:managed}) that may be mutated by a method. 
Every method that may be used for code generation, and also every hand-written method that may have to be checked inside a protection span, will need to be annotated with a mutation summary. Also, every method that is called by a method with a mutation summary must also have a mutation summary. We discuss the effort associated with annotating methods in Section~\ref{sec:adoptability}.


Mutation summaries can be checked by using local information only. We discuss the burden of creating annotations in Section~\ref{sec:adoptability}. There will be cases when developers may want to declare additional mutations beyond what is immediate in the method, in order to allow subclasses to perform these mutations safely, a point that we will return to in Section~\ref{sec:interclass}.

\subsection{Managed fields}
\label{sec:managed}
In addition to mutations of arguments and receivers, mutation summaries may include mutations of fields. In the case of object (reference) fields, except strings, this should be on the basis of what resources have been mutated. In the case of primitive or string fields, we consider them to be mutated if an assignment occurs.

Class designers may partition the fields of a class into \emph{managed fields} and unmanaged ones. Fields are unmanaged by default. Managed fields must occur in mutation summaries whenever they are mutated. They are available, together with method arguments and newly created objects, for use by the planner as part of solutions, and they can be protected by protection spans, since we can track what mutations may occur on them. Unmanaged fields are handled entirely by the programmer and cannot be protected or used in planning. This gives the class designer an encapsulation mechanism, by which the concrete implementation of an abstract resource can be represented as a set of unmanaged fields, which are then guaranteed to be left alone and uninterfered with by subsequent planning attempts. Managed fields are assigned to an abstract resource using the syntax \code{managed(r)}. We give an example in figure \ref{fig:recordSet}. The resource \code{records} is implemented using the fields \code{data} and \code{lastRecord}. Because they are unmanaged, generated code will not refer to them directly, but they can be accessed through the methods \code{addRecord} and \code{getLastRecord}. Subclasses can redefine the \code{records} resource freely, for instance using a \code{LinkedList} instead. The field \code{policy} is managed, and is the subject of a query in the \code{setInverseSorting} method. Every method that mutates the \code{policy} field must now report this, and we are also able to protect its resources.


\begin{figure}
\begin{lstlisting}
class RecordSet {
	Record[] data;
	int lastRecord;
	
	labels(Record) toAdd, added, lastRecord;
	resources records, recordPolicy;
	managed(recordPolicy) unique SortingPolicy policy;
	
	//(initialisers and other code omitted)
	
	void addRecord(Record r) mutates records: 
		r: toAdd, +added {
		data[lastRecord] = r;
		lastRecord++;
		sort();
	}
	
	Record getLastRecord() 
		result: +lastRecord {
		return data[lastRecord];
	}
	
	void setInverseSorting()
	//Alternative: mutates policy
		mutates recordPolicy: { 
		#transform(policy, inversePolicy)
	}
}
\end{lstlisting}
\caption{A record set that tracks some number of records, with a configurable sorting policy. Note that the \code{policy} field is managed, which means that methods must report mutations to it, that it can be protected, and that it can be used as part of query solutions. \code{Unique} indicates that the field is strictly unique (Section~\ref{sec:uniquenessProper}).}
\label{fig:recordSet}
\end{figure}

There are two ways to report the mutation of a field such as \code{policy} in Figure~\ref{fig:recordSet}. If we indicate that a resource of the owning class is mutated, as done in the figure, we allow overriding classes to potentially place new fields in the resource and mutate them. If we indicate specific fields and their resources, we restrict subclasses' overriding ability, but we have more precise information about the method locally, which may allow internal code generation to succeed in some cases where it otherwise would have failed. If mutations of specific private/protected/package fields are declared, they are automatically converted to the corresponding resource from the point of view of outside classes that cannot see the field.

As for communication between managed and unmanaged parts of a class, several policies are possible. We suggest the use of "accessor methods" for use in code generation that read or write unmanaged fields and describe the returned data or the achieved effects in terms of labels and resources. An example of this is the \code{addRecord} and \code{getLastRecord} methods in Figure~\ref{fig:recordSet}. 

\subsection{Mutation summaries and queries}

From the example in Figure~\ref{fig:recordSet} a question arises: what should happen if the solution to the query in \code{setInverseSorting} requires more mutations than what is allowed by the mutation summary? In this case, the method specifies that only the field \code{policy} may be mutated. One design possibility would be to never allow solutions to exceed protection summaries. Another would be to generate code freely and then rewrite the mutation summaries as necessary. In this latter case, the generation of one solution might trigger a cascade of necessary regenerations across a code base, which seems undesirable. We expect that in practice, developers will have some summaries they are willing to change and some that they are not (for instance, internal data such as private fields should be much more plastic), and therefore, a Poplar implementation should be configurable as to what summaries may be rewritten and which ones should be respected when code is generated.

\subsection{Uniqueness}
\label{sec:uniquenessProper}

In order to apply mutation summaries to the checking of a protection span, it is necessary to reason about aliasing of variables. Consider the following example.

\begin{lstlisting}
void m(Socket s1, Socket s2) {
	#transform(s1, type.open) {
		// ...
		s2.close();
		// ...
	}
}
\end{lstlisting}

The intention here is to open the socket \code{s1} and keep it open until the end of the method. In the middle of the span, \code{s2} is closed. Clearly, if s1 and s2 could be the same object, \code{s2.close()} is invalid. This statement can only be permitted if we can know for certain that they cannot be the same. We should conservatively assume that two sockets may be the same if there is any doubt. 

Many sophisticated schemes for reasoning about aliasing have been investigated; many of them revolve around some kind of uniqueness or ownership property. Evaluating sophisticated approaches to be used with Poplar is beyond the scope of this paper, and here we will settle for a simple uniqueness system. The idea exists in various forms in literature, see for instance~\cite{Minsky:1996fk}. Since it is slightly crude, it may constrain the plan search process, but it should be easy to see the intention behind its design and validity. We will annotate references, including method arguments and return values, with the following \emph{uniqueness kinds}.

\begin{description}
\item[Normal] is a reference without any constraints. It permits any number of existing aliases, and any number of aliases may be created in the future. References are normal by default.
\item[Maintain] indicates that the method does not create any aliases of the argument. However, a value being passed in may already have existing aliases. 
\item[Maintain retains], or \code{maintainr} for short, indicates that the method may create a single new alias using a "destructive read". In other words, the caller cannot retain a reference to an argument that has been passed in this way, but must nullify it.
\item[Unique] is a refinement of maintain. No new aliases may be created, and the value must have no preexisting aliases.
\item[Unique retains], or \code{uniquer} for short, is a refinement of maintain retains and of unique. The value must have no aliases, and at most a single new alias may be created.
\end{description}

In addition to method arguments, these also apply to return values. For newly created objects, unless otherwise specified, we may select a uniqueness kind to attribute to them. The idea with this system is that the number of aliases of an object cannot increase as a result of passing it to some method as one of the maintain or unique argument kinds. In addition, receivers of unique arguments can assume that the incoming values are definitely unshared.

Now consider the example given above again. If \code{s1} is unique or uniquer, then clearly \code{s2} cannot be an alias of \code{s1}, so the \code{s2.close} statement is permitted. If \code{s2} is unique or uniquer, then the same is true, so in this case too, \code{s2.close} is permitted. If neither is unique or uniquer, the statement cannot be permitted.

We mentioned destructive reads as a means of transferring references. An example would be as follows.
\begin{lstlisting}
void setX(maintainr Object x) {
	this.x = x;
}
void doSet() {
	Object x = new SpecialX(); //Assuming unshared and permission to assign
	setX(x);
	x = null; //Reference destroyed after being passed on
}
\end{lstlisting}

However, it is not always necessary to nullify references like this after using them in a destructive read. For variables that expire at the end of the method, like this one, it is sufficient to never access them again (at least not in a way that may have side effects or create additional aliases).

Note that even for a maintain or unique argument or return value, we permit the creation of dynamic aliases as part of method invocation or returning a value. For instance, the code shown in Figure~\ref{fig:dynamic} is valid.

\begin{figure}
\begin{lstlisting}
class Container {
	unique Object x; //definitely not aliased, no aliases may be created
	maintain Object getX() { //unique may flow to maintain
		return x; //caller is not permitted to create a heap alias
	}
	
}
//...

int computeHash(Container c) {
	Object x = c.getX(); //creates an alias, but only locally
	return x.hashCode();
}
\end{lstlisting}
\caption{Dynamic aliases may be created for the sake of invoking methods and receiving a return value. The rules of unique and maintain references apply to static aliases only, i.e. the ones that are stored in the heap.}
\label{fig:dynamic}
\end{figure}

\begin{figure}
\begin{center}
\includegraphics[scale=1.1]{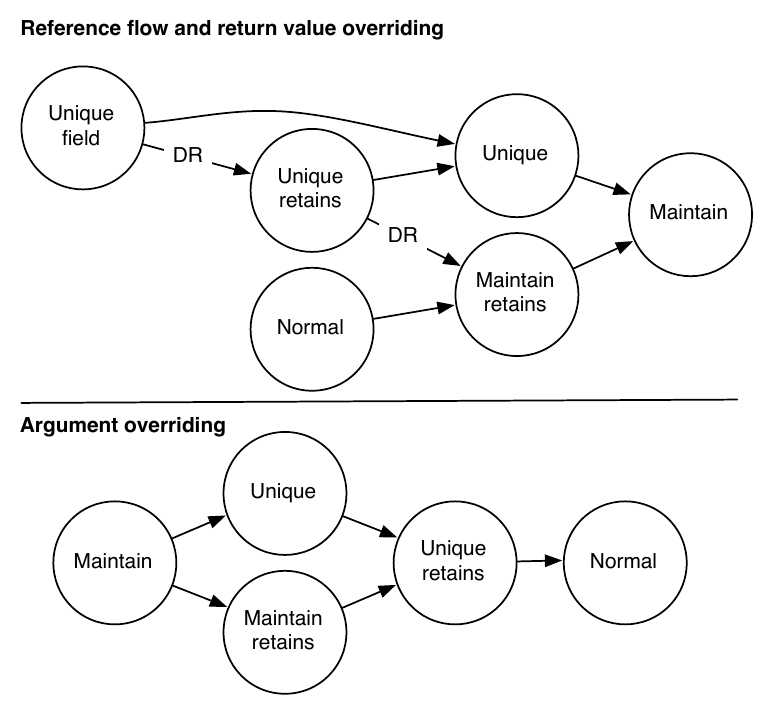}
\caption{\emph{Top:} Permitted argument flow and return value overriding of uniqueness kinds. For instance, a value received as \code{uniquer} can be passed as a \code{maintainr} argument if a destructive read (DR) takes place. Every kind can be passed to itself, with DR as necessary.
\emph{Bottom:} Permitted overriding of argument kinds. \code{Maintain} arguments may override \code{unique} arguments, and so on. Both of these relations are transitive.}
\label{fig:flowOverriding}
\end{center}
\end{figure}

These argument and field annotations can easily be verified syntactically by inspecting local code only. Overriding methods must have compatible uniqueness flags for their arguments and return values. We show permitted overriding for argument kinds and return values in Figure~\ref{fig:flowOverriding}, as well as permitted reference flow. 

We now consider how to apply the uniqueness kinds to mutation summaries. Consider again the previous example:

\begin{lstlisting}
void connectAndSend(maintain Socket s, SocketAddress sa, byte[] data): 
	mutates s.connState, s.data: {
		s.connect(sa);
		s.send(data);
}
\end{lstlisting}

If the argument \code{s} had been normal instead of maintain, the summary would instead have been:
\begin{lstlisting}
mutates any(Socket).connState, any(Socket).data
\end{lstlisting}

This is because we cannot know which object will be mutated in the case of a normal reference. Recall that it is also permitted to pass a normal reference in the place of a maintain argument, for instance in the connectAndSend method given above. In this case, we also have to upgrade the mutation summary so that it is considered to mutate these keys on any object, instead of just on s. When we generate code this kind of transformation should be carried out automatically.

\subsection{Checking protection spans}
When we check the safety of a protection span, we inspect the mutation summary of every statement inside the span, making use of aliasing constraints and summaries as follows. A variable-resource pair is associated with each span. If the variable that is protected is \code{unique}, then it is unshared, and clearly the protected resource is safe unless that variable is passed directly as an argument in such a way that this resource will be violated, according to the summary of the invoked method. If the variable is not \code{unique}, then we must avoid mutations of the form \code{any(T).r}, where \code{T} is a type to which the protected variable is assignable, and \code{r} is a sensitive resource. If we have access to other variables than the protected one, of a compatible type, and that are unique, then we are allowed to violate the sensitive resources on those variables, since they can still not be aliases of the protected variable. We aid this scheme by tracking the potential aliasing state of new variables as we pass them to methods, keeping them maximally unshared when we generate code.


\section{Interclass protocols, subclassing and resource linking}
\label{sec:interclass}
\subsection{Subclassing}
Generally, class elements in an inheritance-polymorphic language such as Java must abide by the substitution principle, that is, we must be able to substitute any concrete subclass for a declared class type without any loss of safety properties or expected functional properties of the program. The various Poplar elements are affected in different ways by this requirement. 

\begin{itemize}
\item Subclassing must weaken preconditions or leave them unchanged, or strengthen postconditions or leave them unchanged. 
\item Mutations must be the same or lesser, and established effects (set up using the \code{[*x]} notation), in so far as they were declared on the superclass level, must be set up in the same resources or in smaller or fewer ones. 
\item Protocols in subclasses must be subprotocols of the overridden protocols. 
\item Uniqueness kinds of method arguments and return values should be overridden in an acceptable way. The applicable constraints are given in Figure~\ref{fig:flowOverriding}.
\item New abstract resources may be added in subclasses as subresources of already defined resources.
\item Existing abstract resources may be completely redefined by subclasses -- the behaviour of their member methods may change, in terms of how they affect unmanaged fields -- since their definition is never exposed. 
\item Subclasses may define new managed fields and add them to existing or new abstract resources. Subclasses cannot remove managed fields from resources they have already been added to, and they cannot associate preexisting fields with preexisting resources.
\end{itemize}

\subsection{A Swing example}

We introduce the remaining features of Poplar using an example based on the Swing GUI toolkit. In this example, we will construct a GUI library supporting efficient application development. The central classes will be \code{SmartFrame} and \code{SmartWidget}. When \code{SmartWidgets} are installed in the \code{SmartFrame}, they supply a Swing component to be displayed, as well as commands that operate on it, which need to be made accessible to the user through the GUI. Depending on the concrete subclass of SmartFrame, the commands may need to be displayed in different ways. 

First, we introduce the \code{SmartWidget} and \code{Command} classes (we have omitted visibility levels and some boilerplate code for brevity):

\begin{lstlisting}

class SmartWidget {
	/* ... */
	List<Command> getCommands() { ... }
	JComponent getComponent() { ... }
}

class Command {
	/* ... */
	String getTitle() { ... }
	int getMnemonic() { ... }
}

\end{lstlisting}

Then, the SmartFrame is as shown in Figure~\ref{fig:smartframe}.


\begin{figure}
\begin{lstlisting}
abstract class SmartFrame {
	resources appearance, widgets, commands;
	managed(appearance) unique JFrame frame = new JFrame();
	managed(appearance) unique JPanel panel = new JPanel(new BorderLayout());
	List<SmartWidget> widgetList = new ArrayList<SmartWidget>();
	
	SmartFrame() { setup(); }
	
	//We anticipate that subclasses may want to mutate panel
	void setup() [!appearance] mutates frame.appearance.size, frame.contents, panel: {
		frame.setSize(500,500);
		frame.add(panel); 
	}
	
	void installWidget(maintainr SmartWidget w) [!widgets] mutates panel.contents, commands, widgetList: {
		panel.add(w.getComponent(), BorderLayout.CENTER);
		installCommands(w.getCommands());		
		widgetList.add(w); 
	}
	
	abstract void installCommand(Command c) [!commands];	
	
	void installCommands(maintain List<Command> commands) [!commands] 
		mutates widgetList: {
		for (Command c: commands) {
			installCommand(c); 
		} 
	}
	
	void show() [!appearance] mutates frame.appearance { frame.setVisible(true); }
	
	//We automatically infer that the return value is linked to this.appearance
	unique JFrame getFrame() {		
		return frame; 
	}
}
\end{lstlisting}
\caption{The \code{SmartFrame} class.}
\label{fig:smartframe}.
\end{figure}
We have defined three resources for the \code{SmartFrame}: \code{appearance}, \code{widgets}, and \code{commands}. Appearance is defined as \code{setup()} and \code{show()}, \code{widgets} is defined as \code{installWidget()}, and commands is defined as 
\code{installCommand()} and\\ \code{installCommands()}. Note that the frame field can be obtained using \code{getFrame()}. The link to the \code{appearance} resource will be automatically inferred, and any external classes that obtain it will be aware, from a code generation point of view, that mutating it would also mutate the \code{SmartFrame} \code{appearance} resource.

From the \code{setup()} method we can also discern that \code{JFrame} declares a resource called \code{JFrame.appearance.size}, and a resource called \code{JFrame.contents}. This kind of subresource mutation propagates upward, also mutating \code{frame.appearance} and the universal root resource \code{frame.root}. \code{Setup()} also declares that it mutates the panel object, even though the method body does not access it. This declaration is added in anticipation of subclasses that may need to mutate it.

We have added the strictest possible uniqueness flags to fields and method arguments. We do not add any particular flags to the argument of \code{installCommand(Command)}, since implementations will need to create several aliases.

\subsection{Concrete subclasses}
We now introduce a concrete subclass of \code{SmartFrame}, \code{MenuFrame}. Its source code is given in Figure~\ref{fig:menuframe}.

\begin{figure}
\begin{lstlisting}
class MenuFrame extends SmartFrame {
	managed(appearance) unique JMenuBar menuBar; 
	managed(appearance) unique JMenu menu;
	void setup() [!Appearance] mutates menuBar.contents, menu.contents, frame.contents: {
		super.setup();
		menuBar = new JMenuBar();
		menu = new JMenu("Commands");
		menuBar.add(menu);
		frame.setJMenuBar(menuBar);
	}	
	void installCommand(Command c) [!commands] mutates menu.contents: {		
		//All code in this method can be replaced by the following query:
		//	#produce(Object, installedInGUI)
		
		JMenuItem item = new JMenuItem(c.getTitle());
		item.setMnemonic(c.getMnemonic());
		item.addActionListener(c);
		menu.add(item);	
	} 
}
\end{lstlisting}
\caption{The \code{MenuFrame} class.}
\label{fig:menuframe}
\end{figure}

\code{MenuFrame} redefines and extends the original appearance and command resources with additional concrete state. Here, the overriding \code{setup()} method is allowed to mutate resources of \code{menuBar} and \code{menu}, even though this was not declared at the original \code{setup()} method in SmartFrame. This is only acceptable because menuBar and menu are newly defined in this class, and because they are strictly unique. If they had not been strictly unique, each \code{setup()} method up to the level of \code{SmartFrame} would necessarily have had to expose mutations such as \code{any(MenuBar).contents}. This requirement could possibly be lifted with a more sophisticated uniqueness/ownership analysis. Because \code{menuBar} and \code{menu} are reported in summaries, they are considered to be managed, and thus we are able to protect effects on them.

The other concrete subclass of \code{SmartFrame} is \code{ToolbarFrame}. It is given in Figure~\ref{fig:toolbarframe}.

\begin{figure}
\begin{lstlisting}
class ToolbarFrame extends SmartFrame {
	managed(appearance) unique JToolBar toolBar;	
	void setup() [!appearance] mutates panel.contents: {
		super.setup();
		toolBar = new JToolBar();
		panel.add(toolBar, BorderLayout.PAGE_START);		
	}
	void installCommand(Command c) [!commands] mutates toolBar.contents: {
	//All code in this method can be replaced by the following query:
		//	#produce(Object, installedInGUI)
		
		JButton b = new JButton(c.getTitle());
		b.setMnemonic(c.getMnemonic());
		toolBar.add(b);
		b.addActionListener(c);
	}
}
\end{lstlisting}
\caption{The \code{ToolbarFrame} class.}
\label{fig:toolbarframe}
\end{figure}

\code{ToolbarFrame} mutates \code{panel.contents}, which is allowed since the original \code{SmartFrame.setup()} anticipated this.  This subclass, too, has extended the definitions of both the commands and the appearance resources. Again, mutating \code{toolBar.contents} is acceptable in \code{installCommand()} since \code{toolBar} is unique and declared in this class.

\subsection{Interclass protocols}
Let us now consider how we can use Poplar to generate code for the task carried out by \code{installCommand} in each of the two concrete classes. Both the toolbar case and the menu bar case require interaction between classes to fulfil their functionality: \code{JButton} and \code{JToolBar} in the former, \code{JMenuItem} and \code{JMenu} in the latter. For such cases, Poplar provides interclass protocols, which describe how different classes may meaningfully interact. For the toolbar and the menu bar case, we may provide two different protocols which have some definitions in common, as shown in Figure~\ref{fig:commandEnhanced}.

\begin{figure}
\begin{lstlisting}
class Command {
	labels(int) actionMnemonic;
	labels(string) actionTitle;
	labels(JButton, JMenuItem) installedInGUI;
	protocols(JButton, JMenuItem) addAction;
 
	int getMnemonic()
		result: actionMnemonic {...}
	String getTitle()
		result: actionTitle {...}

	external JButton(String title)		
		title:actionTitle,
		result: +addAction@1;
	external JButton.setMnemonic(int mnemonic)		
		mnemonic:actionMnemonic,
		this: addAction@1->2
	external JToolBar.add(JButton b)
		b: addAction@2->3
	external JButton.addActionListener(ActionListener al)
		this: addAction@3->4, +installedInGUI
		
	external JMenuItem(String title)
		title:actionTitle,
		result: +addAction@1;
	external JMenuItem.setMnemonic(int mnemonic)
		mnemonic: actionMnemonic,
		this: addAction@1->2
	external JMenuItem.addActionListener(ActionListener al)
		this: addAction@2->3, 
	external JMenu.add(JMenuItem mi)
		mi: addAction@3->4, +installedInGUI		
}
\end{lstlisting}
\caption{Enhanced \code{Command} class with interclass protocols. The \code{installedInGUI} label can be queried for to produce an effect involving multiple interacting classes.}
\label{fig:commandEnhanced}.
\end{figure}

In this case, the protocols have been added inside the Command class, although in principle they could reside in any interface. A best practice should be to place interclass protocols inside the most specific classes that participate in them (i.e. the most dependent), when such a distinction can be made. The numerical state specification ($1 \rightarrow 2$ etc.) is a convenient shorthand when the intermediate states have no useful meaning, and thus no natural names.

Note that we are defining protocols that involve external classes (\code{JButton} and \code{JMenuItem}, respectively). The two protocols have the same name (\code{addAction}) but cannot be interleaved or mixed up, since the state transitions occur on different types. We provide the ``goal label'' \code{installedInGUI} for both of the two types. 

Also note that these "external" annotations cannot be used on their own. They are \emph{overlaid} with the corresponding annotations for these methods inside the JToolBar and JMenuItem classes, so that mutations and other constraints can be taken into account. For external annotations to be valid, they have to be combined with a non-external annotation for the same method.

Given these protocols, in place of the four statements in \code{ToolbarFrame.installCommand}, we could use the following query:

\begin{lstlisting}		
void installCommand(Command c) [!commands] mutates toolBar.contents: {
	#produce(Object, installedInGUI)
}
\end{lstlisting}

The exact same query would also work inside MenuFrame.installCommand, and produce different solutions in the two cases. With this change, the supplier of the Command class may change the definition of the fullyAdded label freely in the future, perhaps providing alternative ways to achieve it or changing the algorithm slightly, requiring other inputs. Also, the code expresses more precisely what the programmer's intention is.

\section{Planning and code generation}
\label{sec:planning}
The design of Poplar does not restrict the choice of planning or search algorithm that is to be used for the code generation, but in early experiments on a prototype, we have found Partial Order Planning (POP) to be a useful algorithm. We now discuss how POP can be applied to finding Poplar solutions.

POP gradually refines a partial ordering of some set of actions. In principle, it searches the space of all possible plans, instead of searching the space of all possible states, as many planners do. POP is a good fit for the problem addressed here, because 1) Java statements are already in some sense partially ordered, through dataflow dependencies for instance, and 2) POP is relatively easy to understand and influence, and it should be a good fit for situations where some amount of human interaction may be needed, for instance in tweaking annotations, disambiguating between values and so on. The basic idea of the POP algorithm is that it gradually strengthens an ordering of actions, inserting causal links (connecting post- and preconditions of related actions) and new actions as necessary while maintaining a set of open preconditions. Conditions will be either of the form \code{new(T, l)}, indicating a new variable of a given type and label, or \code{label(x, l)}, indicating that a variable has a certain label.

A backward search from the goal towards the initial condition is performed as follows.

\begin{itemize}
\item Step 1. Initialise the plan to have two pseudo-actions \emph{start} and \emph{finish}. The effects of the start action are identical to the assumed environment of the plan, i.e. the starting conditions. The preconditions of the finish action are identical to the goals of the plan.
\item Step 2. If there are no open preconditions, stop. A solution has been found.
\item Step 3. Select an open precondition in the current plan.
\item Step 4. For all available actions that achieve the precondition and are either already in the plan or not in the plan, create a successor plan with this action added. Also add ordering constraints and causality links (which pair preconditions with postconditions) for the new action.
\item Step 5. For all the successors, resolve any conflicts among the causality links that might have arisen by strengthening the ordering constraints. If this is not possible, discard the successor.
\item Step 6. Recurse on each successor plan. Go to step 2.

\end{itemize}
In addition to the plan itself, the planner needs to keep track of the following mappings:

\begin{itemize}
\item The planning context, which is the set of all variables that are available to the planner at any given point. Inside a method body, it is initialised to contain the arguments of the method, managed fields, and all local variables that have been initialised prior to the query expression.
\item type($x$), which maps each variable in C to its type and uniqueness flag.
\item labels($x$), which maps each variable in C to its label set.
\item resources($x$, $l$), which maps a variable and label pair to the set of abstract resources of that variable which the label may reside in.
\item uniqueness($x$): the uniqueness kind of each variable. 
\end{itemize}

\subsection{Variable reuse and progress}
When the planning algorithm needs to fulfil a precondition, before attempting to construct a new value, it first looks among the known existing values, to see if any of them fits the requirement. We consider two variables with the same type and the same labels to be equivalent, and if one has labels that are a superset of the labels of another, but the types are equal, then the variable with fewer labels is redundant. If 

\[
\code{type}(x) <: \code{type}(y) \wedge \code{labels}(x) \supseteq \code{labels}(y)
\]

 where $<:$ denotes subtyping, then generally, $x$ can take the place of $y$. 

More precisely, a new variable is useful if it has at least one of the following:
\begin{enumerate}
\item A new type which was not previously available for use
\item A new type/label combination which was not previously available for use
\item A stronger uniqueness flag than what was previously available for some type/label combination. 
\item An effect that resides in a smaller set of resources, or a new set of resources, for some type/label combination.
\end{enumerate}

From this view we obtain the important ability to tell when a plan search is making no further progress. Given a finite number of actions, and a finite number of labels (which will always be the case), we gradually make progress from having produced nothing to having produced all possible useful values. Furthermore, if we remember the state of the plan after each search step, we can also identify when the search is cyclically creating and destroying the same conditions, making no progress. So the plan search is decidable, although in practice we will want heuristics to avoid the full exploration of this very large search space.

\subsection{Heuristics}
 The POP algorithm needs at least two heuristics: one for selecting the next open precondition to plan for, and one for selecting the most appropriate action to attempt for a given precondition, in the event that several are available. Designing appropriate heuristics is in the scope of future work. In our early prototype, we have attempted to use the following naive  precondition heuristic with some success:
 
\begin{itemize}
\item If a precondition has no available actions, select it (in order to fail quickly and backtrack).
\item If a precondition has only one available action, select it.
\item Otherwise, select the precondition with the smallest amount of available actions, in an attempt to lock in necessary decisions early and reduce the size of the search space.
\end{itemize}

For the action heuristic, we have experimented with schemes that favour syntactic locality when selecting actions. For example, prefer fields in the same class over invoking a method, prefer local methods over methods in other classes, and prefer classes in the same package over classes in other package.

\subsection{Forced inclusion}
In the \code{installedInGUI} example in the previous section, there are two ways to produce the desired effect. One makes use of classes such as JMenuItem and JMenuBar, the other makes use of JButton and JToolBar. As long as the planner gives all possibilities the same amount of search effort, it should find the appropriate solution in each of the two client classes. However, there may be situations when one wants to force a particular solution or influence the planner to do something that does not correspond to the "best fit" in a particular context. For this, we introduce the \code{with} syntax, so that one can write a query such as the following.
\begin{lstlisting}
#produce(Object, installedInGUI) with JMenuItem
\end{lstlisting}
This kind of modifier is the dual of a protection span. Protection spans identify resources that are not to be used, and with-clauses identify classes or labels that must be used as part of the solution, and get the highest priority in case of ambiguity. This may help reassure programmers who are naturally hesitant to use tools that generate code which may vary from time to time.

\section{Discussion and practical concerns}
\label{sec:discussion}

\subsection{Verifying upgrades}

 When published classes change, their interfaces are re-published if protocols or method annotations have changed. Checking whether an existing integration is still valid is then a matter of checking whether the new annotations are compatible with the old ones. 
The Poplar code generation tool should output a set of \emph{integration assumptions} for each query that it solves. They would capture the full signature of each method or field assumed in each plan, and be stored together with the Java class files. For a future upgrade to be compatible without code regeneration, we would compare future method and field signatures with what is stored in the assumptions; every step should comply with the same rules as for subclassing (see Section~\ref{sec:interclass}).

%

Using such integration assumptions, when we check the validity of an existing integration, we may also be able to identify minimal parts of the existing integration plans that are invalid, and reconstruct only those, leaving the remaining parts of the integration intact. 

\subsection{Applicability and limitations}
In the introduction, we mentioned several kinds of breaking changes that may occur in Java source code, including syntactic changes, protocol changes, semantic changes, and so on. We believe that Poplar would be able to address many of these issues automatically. 

An important limitation of Poplar is that control flow constructs are not generated. For instance, we do not generate loops or if-statements. This means that when flow control is required, as in the case of an iterator which needs to repeatedly test a truth condition in a loop, this information should be communicated externally. For an iterator, separate queries could be used for the truth condition and for the loop body.

Poplar assumes, in \code{\#produce} queries, that the type of the desired value is known. The more specific this type is, the easier the query will be to resolve. However, this may also couple the client strongly to a particular implementation, in the case where the same functionality is supplied by different components that have no shared type hierarchy.

Exceptions have not been considered in this paper. Exception handlers can play a role similar to protection spans, effectively constraining the methods that can be considered for use in solutions, but further work is needed to clarify the details.

\subsection{Reliability and predictability}
Developers that use Poplar might be concerned about the fact that generated code can vary from time to time. Our basic means of ensuring sensible outcomes of code generation are as follows. Firstly, variables with the same type and labels must be truly equivalent. If they are not substitutable for each other, more labels should be added to disambiguate. Second, protection spans should be used to protect those resources that may not be mutated. In the case of handwritten code that surrounds queries, developers may need to identify resources that need to be protected manually.

The outcome of a code generation attempt is dictated by the choices of algorithm and heuristics. In general, we think that a good algorithm should strive to favour short valid solutions over longer valid solutions, once other needs have been taken into account. Thus, one situation that may lead to unexpected outcomes is if a component supplier makes it \emph{easier than before} to produce a given type/label type, thereby making the altered protocol a simpler means of production than existing protocols. In this case, at the next regeneration, the shortened code fragment may out-compete some existing solutions in the quest to be shortest, which could have adverse consequences if annotations are not precise enough. 

The \code{with} modifier, which forces the inclusion of an element in a plan, as well as the ability to decide which variables are managed (available for planning) and unmanaged should be valuable countermeasures towards unwanted outcomes.

\subsection{Adoptability}
\label{sec:adoptability}

In adopting Poplar for use in an existing Java code base, it is necessary to add annotations concerning mutations, uniqueness, protocols, resources and so on. We have seen that information such as mutation summaries and uniqueness can easily be inferred locally on a class by class basis, and such an inference tool should be simple to make. In addition, if one sets up a temporary mapping between unmanaged fields and abstract resources in a class (to be discarded afterwards), it should be easy to infer a partial set of resource definitions (\code{[!x]} annotations). It is even possible to infer protocols, as in~\cite{Mandelin}.
In addition, when Poplar is introduced into an existing code base, it should be possible to start with a small number of queries and annotations, and gradually expand the use of Poplar within the code base.
%


%




\section{Related work}	
\label{sec:related}
\subsection{General}
The notion of component-based software was discussed as early as 1965 in a study by McIlroy~\cite{Il68}. 

Shaw~\cite{Shaw1993} carried out one of the first studies that highlight the brittleness of procedure call based component interconnections in 1993.

Aspect-oriented programming~\cite{aspectj} (AOP) weaves together code from different aspects and inserts them, at compile time, in locations that are specified declaratively. The acceptance of AOP indicates that developers are willing to use tools that generate extra code at compile time, provided that they can easily understand what will result.

One of the first descriptions of the partial order planning algorithm was given by Allister and Rosenblatt~\cite{Mcallester1991}. 

\subsection{Code synthesis}
Two of the most fundamental approaches to code synthesis are deductive~\cite{136556} and inductive\cite{simpleInductive} synthesis.

Haack~\cite{Haack2002} has created a system that unifies software components in the ML language by generating code from uninterpreted, atomic annotations, much like Poplar. However, given the differences between ML and Java, and that ML unification is the central technique in Haack's work, it is not clear if the findings can be applied to an imperative object-oriented setting.

Bastani et al.~\cite{Bastani2006} describe glue code synthesis for embedded systems based on code patterns. A code pattern is a code fragment with pre- and postconditions expressed in OCL, and thus it is not integrated with the programming language. Synthesis is carried out by composing code patterns according to predefined rules when possible. The system is dependent on a general theorem prover.

Ireland and Stark~\cite{Ireland2006} have designed a system that combines proof plans and partial order planning to generate small imperative programs. The examples given are mathematical routines specified using Hoare logic for pre- and postconditions. The heuristics and proof critics used here are adopted from the literature on Structured Programming, which describes principles that are to be used for manual goal-directed programming. This should be a valuable avenue for future investigation of heuristics for Poplar.

Jha, Gulwani et al.~\cite{oracleGuided} have described a system that synthesises loop-free imperative programs from components using an SMT constraint solver. However, ``components'' here are not in the CBSD sense, but rather, they are algorithm fragments. The specification language used is fine grained, and the computational power is spent on a much more fine grained level than in Poplar.

\subsection{Evolution and adaptation}
Dig~\cite{Dig2005} has carried out an empirical study of in component evolution. Strikingly, he found that between 81\% and 100\% of all breaking changes in several large systems were due to refactorings. Vasa et al~\cite{Vasa2007} studied the evolution of several large Java libraries in terms of the proportion of derived types and the distribution of classes into various layers of the architecture. They find that in quantitative terms, interfaces do not grow very much over time. However, in light of the findings from~\cite{Dig2005}, it is likely that many subtle changes that their quantitative approach cannot capture have occurred, such as protocol changes and semantic changes.

Zaremski and Wing~\cite{zaremskiWing} have studied component discovery and matching. They classify several kinds of matching relations, including plug-in matches, exact matches, matches with slightly weaker or stronger specifications than the query given, and so on. It is possible that their classification of different matches could be applied in a future Poplar integration algorithm.

Becker et al~\cite{Becker2006} discuss an engineering approach to adaptation. They give a taxonomy of possible component mismatches, ranging from signatures, assertions and protocols to quality attributes and semantic concepts. They also discuss the use of design patterns to overcome these mismatches. Design patterns can provide good leverage in terms of reducing the problems introduced by evolution, but cannot remove such problems since they are bound by the programming language.

Kell~\cite{Kell2009,Kell2007} focuses on adaptation on a binary, operating system level (shared objects in Unix). He identifies two myths about component based software development: that there is a notion of a "matching module" that can be discovered and plugged in, and that development of components occurs in order: depended-on first, dependent later.

\subsection{Java effect systems}
Boyland and Greenhouse~\cite{objOrientedEff} describe an effect system based on \emph{abstract regions}, which are very similar to the abstract resources used in Poplar. However, the abstract regions are sets of concrete fields, instead of being sets of methods, and the intention is to capture reads and writes, rather than mutation and residence, as in our case. They also introduced the idea of effect summaries based on the abstract regions. By generalising beyond field reading and writing, focussing instead on method invocations, Poplar can define resources as corresponding to practically any effects that can be achieved with Java code, including effects achieved with native methods. Boyland and Greenhouse's analysis was later adapted to Middleweight Java -- a core imperative fragment of Java -- by Parkinson et al~\cite{cite-keymj}, and this version was formally proven to be correct. \emph{Data groups}, by Leino et al~\cite{Leino:1998fk} are a slightly simpler idea that predates abstract regions.

Pearce~\cite{jpure} has introduced a modular purity system for Java. It makes judgments based on the locality and freshness of variables. Unlike in Poplar, where effects are identified in terms of abstract resources, Pearce's system identifies the absence of observable heap (field) writes.

\subsection{Typestate checking}
Typestate was first introduced by Strom and Yemini~\cite{Strom1986} as an approach to checking valid interactions with primitive types and simple data structures.
 
Deline and F\"ahndrich~\cite{Deline} describe how typestate checking can be applied to object-oriented languages. They use a core subset of C\# as an example. A key concept in their system, Fugue, is the notion of \emph{sliding methods}: subclasses may refine the meaning of state transitions, and every overriding method may implement a slightly stronger state change. A sliding method first invokes its corresponding superclass method before implementing its own part of a change. 
Bierhoff et al.~\cite{Bierhoff2007} describe a modular approach to typestate checking that is based on linear logic for interface descriptions. This enables them to track the multiplicity of outstanding references, for instance the number of iterators that have been produced from a collection. Their approach allows protocols to have subprotocols, which has similarities with resources and subresources used in Poplar.

Aldrich et al.~\cite{Aldrich} have described typestate oriented programming, an effort that is being realised in their programming language Plaid. Plaid integrates typestate as a first class language construct; however it is used strictly for checking and safety constraints, and not as an integration aid.
%

Mandelin et al.~\cite{Mandelin} describe a tool that mines typestate-like protocols from code bases presumed to be valid, something that could be useful in generating initial Poplar annotations. Their work was an important inspiration for Poplar. However, the tool is designed for interactive use only, and does not reason about side effects or aliasing. Poplar aims to support fully automated integration and verification of integrations at compile time.


\section{Conclusion and future work}
\label{sec:conclusion}
We have presented Poplar, a Java extension designed for automated component integration based on queries and code generation. Through the use of abstract resources, protocol and label annotations, aliasing constraints and a planning algorithm, Poplar enables controlled generation of integrating code with constrained side effects. By expressing more functional properties and constraints in interfaces than what is usual in object-oriented languages, we describe components in such a way that they can be integrated and re-integrated in a dynamic, flexible fashion. Many API changes and mutations should be compensated for automatically as they occur. We believe that the integration approach presented here could be an important step towards much greater reusability and ease of maintenance in component-based development. 

As for future work, most importantly, we aim to provide a practical implementation and investigate how Poplar works in practice. We also expect to need to investigate more sophisticated aliasing policies, exception handling, effect systems and concurrency support. It is interesting to note that the aliasing policy and the code generation mechanism mutually influence each other: since we do not generate all valid forms of Java code, but only statements with restricted forms, the alias analysis does not need to deal with all possible cases. On the other hand, the more powerful the analysis is, the more freely code may be generated without having adverse effects. 

An essential limitation in the design presented here is the protection spans. By design, they assume that regions that need to be protected are continuous and part of the same method body. However, in practice, effects are often created and destroyed in quite separate locations of programs. Investigating support for this is in the scope of future work.



\acks

The authors would like to thank Liyang Hu, Zhenjiang Hu, Atsushi Igarashi, Michael Nett and the Honiden laboratory's \code{MALACS} group for helpful comments on this work.


\def\urlprefix{}
\def\url#1{}

\bibliographystyle{customabbrvnat}
\softraggedright
\bibliography{bibliography}




\end{document}